\documentclass[12pt,preprint]{aastex}

\usepackage{emulateapj5}

\def\hii{H\,{\sc ii}}
  
\def\kms{\relax \ifmmode {\,\rm km~s}^{-1}\else \,km~s$^{-1}$\fi}
\def\cm-3{\relax \ifmmode {\,\rm cm}^{-3}\else \,cm$^{-3}$\fi}
\def\Jb{\relax \ifmmode {\,\rm Jy\,beam}^{-1}\else \,Jy\,beam$^{-1}$\fi}
\def\mJb{\relax \ifmmode {\,\rm mJy\,beam}^{-1}\else \,mJy\,beam$^{-1}$\fi}
\def\deg{\hbox{$^\circ$}}
\def\arcmin{\hbox{$^\prime$}}
\def\arcsec{\hbox{$^{\prime\prime}$}}
\def\secd#1.#2{ #1\farcs#2 }               
\def\e{$\pm$}
\def\x{$\times$}

\begin{document}

\title{Interferometric Mapping of Magnetic Fields in Star-forming Regions I. W51 e1/e2 Molecular Cores}
\author{Shih-Ping Lai, Richard M. Crutcher, 
Jos\'e M. Girart\altaffilmark{1}, and Ramprasad Rao\altaffilmark{2}}
\affil{Astronomy Department, University of Illinois, 1002 W. Green Street, Urbana, IL 61801; \\
slai@astro.uiuc.edu, crutcher@astro.uiuc.edu, jgirart@am.ub.es, ramp@oddjob.uchicago.edu}

\altaffiltext{1}{Current address: Departament d'Astronomia i Meteorologia, Universitat de Barcelona, 08028 Barcelona, Catalunya, Spain}
\altaffiltext{2}{Current address: Department of Physics, University of Chicago}

\begin{abstract}
We present the first interferometric polarization map of the W51 e1/e2 
molecular cores obtained with the BIMA array at 1.3 mm wavelength
with approximately 3\arcsec\ resolution.   
The polarization angle varies smoothly across the double 
cores with an average position angle of 23\deg\e5\deg\ for W51 e1 and 
15\deg\e7\deg\ for W51 e2.  The inferred magnetic field direction is 
parallel to the minor axis of the double cores, which is consistent with 
the theoretical picture that clouds collapse along the field lines.
However, the magnetic field may not determine the axis of angular momentum 
of these two cores as the field directions of the two cores
significantly differ with the previously measured directions of 
rotational axes.
The polarization percentage decreases toward regions with high intensity, 
suggesting that the dust alignment efficiency decreases toward high density 
regions.  The field directions are highly ordered, and the small dispersion
of the polarization angles implies that magnetic fields are strong 
($\gtrsim$ 1 mG) and perhaps dominate turbulence in W51 e1/e2. 

\end{abstract}



\section{Introduction}

Star formation in molecular clouds is one of the most fundamental
problems in astrophysics.   It has become increasingly evident that 
star formation cannot be fully understood without considering magnetic
fields (see recent reviews by Mouschovias \& Ciolek 1999, and Shu et 
al.\ 1999).    Magnetic fields provide explanations for 
the support of clouds against self-gravity, the formation
and evolution of cloud cores, the origin of supersonic line widths,
the low specific angular momentum of cloud cores and stars,
and the formation of bipolar outflows.
Unfortunately, the magnetic field remains the most poorly measured 
quantity in the star formation process.
In order to advance our understanding of star formation,
it is essential to improve our empirical knowledge of 
magnetic fields in molecular clouds.

The most promising technique to probe magnetic field morphology
in dense molecular cores is to measure the linear polarization
of the thermal emission from spinning, magnetically aligned dust grains
(Heiles et al.\ 1993).   Interstellar dust grains are elongated with 
their minor axes aligned with the magnetic field,
thus the magnetic field direction in the plane of sky is perpendicular 
to the direction of polarization (Davis \& Greenstein 1951; Roberge 1996).
Millimeter interferometers detect the dust emission
in the low optical depth regime with high angular resolution,
thereby revealing the field morphology in the dense cores.  
In the past few years, linear polarization observations with 
the Berkeley-Illinois-Maryland Array (BIMA) of Orion-KL
and NGC\,1333 IRAS\,4A have demonstrated the capability 
and reliability of polarization observations with this telescope
(Rao et al.\ 1998; Girart et al.\ 1999).
These results are not only consistent with the previous 
single-dish observations, but also provide new information 
at resolutions up to 3\arcsec.   Therefore, it is reasonable for us 
to carry out a large linear polarization survey of star-forming cores 
with BIMA.

W51 is a large molecular cloud/\hii\ region complex at a distance 
of 7.0$\pm$1.5 kpc (Genzel et al.\ 1981).  Associated with ultracompact 
(UC) \hii\ regions (Scott 1978), strong infrared and submillimeter continuum 
emission (Thronson \& Harper 1979; Jaffe, Becklin, \& Hildebrand 1984), 
molecular gas (Ho, Genzel, \& Das 1983) and H$_2$O and OH masers
(Genzel et al.\ 1981; Gaume \& Mutel 1987),  
W51 is one of the most active high-mass star formation regions 
in our Galaxy.   W51 e1 and W51 e2 are molecular cores located in the 
eastern edge of W51.  The mass of W51 e1 and e2 are $\sim$ 150 and 
110 $M_{\odot}$, derived from NH$_3$ observations (Zhang \& Ho 1997).
With infall motions identified, these molecular cores are 
in the process of forming OB associations (Ho \& Young 1996).
Subarcsecond VLA observations reveal that W51 e1 contains 
three UC \hii\ regions, e1, e3, e4 (Gaume, Johnston, \& Wilson 1993),
and a possibly dust source, e8 (Zhang \& Ho 1997).
In the vicinity of the UC \hii\ region e2,
there is a compact water maser concentration known as W51 Main
(Genzel et al.\ 1981).
Previous single-dish measurements with resolution $\sim$30\arcsec\
show very low polarization in W51 e1/e2
consistent with non-detection (Kane et al.\ 1993; Dotson at al.\ 2000);
however, this could be caused by averaging over the large beams.
Our observations with 10 times better resolution
will allow us to explore the field structure in the W51 e1/e2 cores.

In this paper, we present the first interferometric polarization
map of the W51 e1/e2 cores at 1.3 millimeter wavelength.
Other sources in our survey will be reported in later papers.

\section{Observations and Data Reduction}

Several observations were carried out from 1999 August to 2000 April 
using nine BIMA antennas with 1-mm SIS receivers and quarter-wave plates.
The digital correlator was setup to observe the continuum emission
with a 750 MHz window centered at 226.9 GHz in the lower sideband 
and a 700 MHz window centered at 230.9 GHz in the upper sideband.   
Strong CO $J$=2--1 line emission was isolated in 
an additional 50 MHz window in the upper sideband.
The primary beam is $\sim$ 50\arcsec\ at 1.3 mm wavelength.
Data were obtained in the B, C and D array configurations 
with projected baseline ranges from 6--170 kilowavelengths, 
resulting in a linear polarization map of the W51 e1/e2 cores 
with a long integration time, $\sim 25$~hours.

The BIMA polarimeter and the calibration procedure are described 
in detail by Rao et al.\ (in preparation).
A pair of quarter-wave plates was placed in front of the linearly
polarized receiver of each antenna in order to enable
the detection of the left ($L$) and right ($R$) circularly 
polarized radiation.   Four cross correlations, 
$LL,\ RR,\ LR$, and $RL$, must be measured at each baseline 
in order to obtain the four Stokes parameters, $I$, $Q$, $U$, and $V$
(Thompson, Moran, \& Swenson 1986). 
Because there was only one receiver at each BIMA antenna,
the quarter-wave plates were switched on a time scale shorter 
than the $uv$-cell transit time to achieved quasi-simultaneous 
dual polarization observations.
Walsh function switching patterns (Thompson, Moran, \& Swenson 1986; 
Harmuth 1969) were used to maximize the efficiency of obtaining four cross 
correlations (Rao 1999).
Sixteen switch patterns were needed to complete a cycle.
The integration time of each pattern was set to 11.5 seconds 
and the plate switch took an additional 2--3 seconds;
therefore, one cycle was about 4 minutes.

Calibration and data reduction were carried out using 
the MIRIAD software (Sault, Teuben, \& Wright 1995).  
The instrumental polarization response (``leakage")
was calibrated by observing the strongly polarized quasars 
3C273 or 3C279 over a wide hour angle range, typically
more than 5 hours.   The observed polarization, including 
contributions from the instrumental polarization
and the polarization of the calibrator, varied with hour angle 
due to the fact that the polarization vector of the calibrator 
rotates with respect to the linear feed horn.   
On the other hand, the instrumental polarization is constant with time.
Therefore, the leakages can be solved by assuming that the
polarization of the calibrator was constant in the frame of sky 
over the whole track.
Task GPCAL of MIRIAD was used to solve simultaneously for the leakages 
and antenna gains from the polarization calibrator.
The average leakage of each antenna was 5.9\%\e0.4\% 
for our observations.  The leakage correction of each antenna was 
applied to visibility data before further processing.

Channels with line emission were carefully flagged out in the visibility 
data of the continuum bands.  
As the continuum emission of the the W51 e1/e2 cores was stronger than 
the phase calibrator (QSO 1751+096), self-calibration was performed 
for refining antenna gain solutions.  The Stokes $I$ image was made 
with Briggs' robust weighting of 0.5 (Briggs 1995; Sault \& Killeen 1998)
to acquire a smaller beam size (2\farcs7\x2\farcs0, PA=1\deg)
for better determination of the self-calibration model.  
The model was used to calibrate the visibility data in order to 
obtain new gain solutions with shorter time intervals.  
The above iterations were repeated until the Stokes $I$ image 
was not sufficiently improved.
The Stokes $Q$ and $U$ images were then made with natural weighting 
to obtain the highest S/N ratio. The resulting synthesized beam
was 3\farcs2$\times$2\farcs3 with PA=2\deg.
Maps of Stokes $Q$ and $U$ were deconvolved and binned to approximately
half-beamwidth per pixel (1\farcs5\x1\farcs2) to reduce oversampling
in our statistics.   We then combined the maps to obtain the observed 
linearly polarized intensity,
\begin{equation}
I_{p,obs}= \sqrt{Q^2+U^2},
\end{equation}
\noindent and the polarization position angle,
\begin{equation}
\phi = 0.5~\tan^{-1}{(U/Q)}.
\end{equation}
\noindent As the polarized intensity is an intrinsically positive 
quantity, Eq.\ (1) will tend to overestimate the polarized intensity.
Therefore, a bias correction for the polarized intensity must 
be performed (Leahy 1989),
\begin{equation}
I_p = \sqrt{I_{p,obs}^2-\sigma_{I_p}^2},
\end{equation}
\noindent where the rms noise value of polarized intensity, $\sigma_{I_p}$,
was taken as the average value of the rms noise in the $Q$ and $U$ maps,
as the noise levels in these two maps are comparable.
The polarization percentage was calculated from
\begin{equation}
p=\frac{I_p}{I}.
\end{equation}
The measurement uncertainty in the position angle,
\begin{equation}
\sigma_{\phi} = 0.5~\tan^{-1}{(\sigma_{I_p}/I_p)},
\end{equation}
\noindent depends on both $\sigma_{I_p}$ and $I_p$, thus it varies across 
the map.  When weighted with $I_p$, the average measurement uncertainty 
in the position angle for our observations was 5\fdg6\e1\fdg7.

\section{Results and Analysis}

Figure 1 displays a contour map of the 1.3-mm continuum from the W51 e1/e2 
cores overlaid with polarization vectors.  
Polarization is called detected at positions where the linearly polarized 
intensity is greater than 3$\sigma_{I_p}$ (1$\sigma_{I_p}$ = 4.7 \mJb) 
and the total intensity is greater than 5$\sigma_I$ 
(1$\sigma_I$ = 27 \mJb, which is dominated 
by incomplete deconvolution rather than thermal noise).    
Under these criteria, the polarized emission extends over an area of
$\sim$12 beam sizes.  
Table 1 lists the polarization measurements in the W51 e1/e2 cores 
at selected positions, separated by approximately the synthesized beamsize.
The distributions of the polarization angle and the polarization percentage 
are shown in Figure 2 and 3, respectively.

\subsection{Continuum and Polarized Emission}

The south and north continuum condensations of the double 
cores in Figure 1 correspond to W51 e1 and W51 e2, respectively. 
W51 e1 is extended (FWHM=5\farcs0\x3\farcs0, PA=11\deg)
and its integrated flux is 8.2 Jy.    
W51 e2 is also resolved (FWHM=4\farcs4\x3\farcs3, PA=$-$8\deg)
with an integrated flux of 13.6 Jy.
We will refer to the intensity maxima of these two cores as 
``e1 1mm peak'' and ``e2 1mm peak''.
Analysis of the continuum flux at multiple wavelengths has shown that 
W51 e1 and W51 e2 contain both free-free emission and dust emission 
(Rudolph et al.\ 1990).  The expected flux of the free-free emission 
at 1.3 mm wavelength for these two cores can be estimated
from their SEDs (Rudolph et al.\ 1990), which are $\sim$ 0.5 Jy
for both cores.  Therefore, the average fractions 
of the free-free emission in W51 e1 and e2 are $\sim$6\% and $\sim$4\%.

The polarized emission does not show two distinct peaks associated with 
W51 e1 and e2, but it mainly arises from an extended region across 
the two cores.   The peak of polarized intensity is at 1\farcs2 south of 
e1 1mm peak (Table 1).  There is also a compact region of polarized 
emission $\sim$4\arcsec\ northwest of e2 1mm peak 
(for convenience, we name this position ``e2 pol NW"; cf.\ Fig.\ 1).
The polarized flux drops to zero between e2 1mm peak and
e2 pol NW, and e2 pol NW shows very different polarization angle.
Therefore, the polarization in e2 pol NW may sample a different region,
and is excluded in our statistical analysis.

Most of the polarized flux is associated with W51 e1.
The elongated shape of W51 e1 is due to the fact 
that it is composed of emission from nearby UC \hii\ regions 
(e1, e3 and e4), a newly discovered continuum source 
(e8: Zhang \& Ho 1997), and the dust emission in the W51 e1 core.
The nearest source to e1 1mm peak is e8, which indicates that e8 is 
stronger than the UC \hii\ region e1 at 1.3 mm.  Because the UC \hii\ 
region e1 is stronger than e8 at 1.3 cm (Zhang \& Ho 1997) and e8 is not 
detected at 2 cm (Gaume et al.\ 1993), the spectral index of e8 is very 
different from free-free emission.  This fact is consistent with Zhang 
\& Ho's suggestion that e8 is a dust-dominated continuum source.  Hence, 
e8 is the most likely site for active star formation in the W51 e1 core.

The position of the UC \hii\ region e2 (accurate to 0\farcs2; 
Gaume et al.\ 1993) is offset $\sim$1\arcsec\ to northwest of e2 1mm peak .
It is known that the molecular gas in W51 e2 is not evenly 
distributed around the \hii\ region (Zhang \& Ho 1997).  
If we consider that molecular gas is in general associated with 
dust grains, then, although the offset is small, it is evident that 
the free-free emission and the dust emission in W51 e2 do not coincide.  
Since the polarized intensity only originates from the dust emission
and the fraction of the free-free emission could be significantly higher 
than the average value at the position of the UC \hii\ region,
this offset may be part of the reason for the absence of polarization 
between e2 1mm peak and e2 pol NW. 
It is interesting that the location of W51 Main is right in 
the polarization gap between e2 1mm peak and e2 pol NW,
and a stream of water masers in W51 Main has been detected
with proper motions headed toward position angle 200\deg\ 
(Lepp\"anen, Liljestr\"om, \& Diamond 1998),
which seems well matched with the direction of the polarization gap.

\subsection{Polarization Angle Distribution}

The polarization angle distribution in the W51 e1/e2 cores
appears to be well matched to the morphology of the cores.
Except that e2 pol NW has a very different position angle 
at 58\deg$\pm$8\deg,
the rest of the polarization vectors are uniformly distributed 
along the major axis of the double cores with an average position 
angle of 21\deg$\pm$6\deg.  
Polarization vectors seem to be approximately parallel to local core edges: 
in W51 e1, polarization angles decrease westward and increase eastward 
from the central ridge to the edge; in W51 e2, position angles 
are close to 0\deg\ at the northeast edge; even for e2 pol NW, 
the difference 
between the polarization angle and the edge is less than 20\deg.
The polarization angles also follow the ridge connecting W51 e1 and W51 e2.
Excluding e2 pol NW, we calculate the mean polarization angles for W51 e1 
and W51 e2 weighted by the measurement uncertainty, which are
23\deg$\pm$5\deg\ for W51 e1 and 15\deg$\pm$7\deg\ for W51 e2. 
Compared to the measurement uncertainty,
the variation of polarization angles corresponding to 
the core morphology is only slightly suggested by our data.

Figure 2 shows the distribution of the polarization
angles in W51 e1 and W51 e2.  The observed polarization angle dispersion 
$\delta\phi_{obs}$, defined to be the standard deviation of 
the position angles, is 4\fdg8\e2\fdg0 in W51 e1 and 6\fdg6\e3\fdg0
in W51 e2 (excluding e2 pol NW).    Interestingly, these values are 
equal to the measurement uncertainty of the position angles $\sigma_{\phi}$, 
4\fdg9\e1\fdg4 in W51 e1 and 6\fdg7\e3\fdg4 in W51 e2.  
The observed dispersion is made up of contributions from the measurement 
uncertainty and the intrinsic dispersion $\delta\phi$, 
$\delta\phi_{obs}^2 = \delta\phi^2+\sigma_{\phi}^2$,
which indicates that the intrinsic dispersions in W51 e1 and e2 
are very small. 
Therefore, if we take $\sigma_{\phi}$ as small as possible
($\sigma_{\phi}-1\sigma_{\sigma_{\phi}}$) and $\delta\phi_{obs}$
as large as possible ($\delta\phi_{obs}+1\sigma_{\delta\phi_{obs}}$),
we can obtain the upper limit of the intrinsic dispersion,
which is 5\fdg8 in W51 e1 and 9\fdg0 in W51 e2.
Our results for the dispersion analysis are summarized in Table 2.

\subsection{Polarization Percentage Distribution}

The average polarization percentage of the W51 e1/e2 cores is 
1.8\%\e0.4\%, including positions for which polarization vectors
are not plotted in Fig.\ 1 because $I_p$ is too small. 
W51 e2 has a lower polarization percentage (1.1\%\e0.3\%) 
than that of W51 e1 (3.2\%\e0.6\%). To better compare with previous 
observations, we convolved our map to 30\arcsec\ resolution with 
a Gaussian beam and obtained 1.3\%$\pm$0.2\% polarization with a position 
angle of 25\deg$\pm$3\deg. The polarization detections we have obtained 
are clearly different from the single-dish results: at 1.3 mm (HPBW=30\arcsec),
Kane et al.\ (1993) measured a polarization percentage of 0.50\%\e0.21\% 
at an angle of $-$31\deg\e11\deg; at 100 $\mu$m (HPBW=35\arcsec),
Dotson at al.\ (2000) obtained zero polarization (0.54\%$\pm$0.58\%).
However, all results could be still consistent, 
because the low S/N ratio of the these single-dish measurements can 
lead to large uncertainty in the polarization angle.
Our observations demonstrate that high resolution is important 
for successful detection and mapping of magnetic fields in molecular cores.

Figure 3 shows plots of the polarization percentage 
versus Stokes $I$.  These plots show that the polarization 
percentage decreases as the total intensity increases in both W51 e1 
and W51 e2.  Such a decrease in the polarization toward high intensity 
regions ({\it the depolarization}) has been commonly seen in polarization 
observations, e.g., L1755 (Lazarian, Goodman, \& Myers 1997), 
OMC-1 (Schleuning 1998), and OMC-3 (Matthews \& Wilson 2000).  
Lazarian, Goodman, \& Myers (1997) have shown that the depolarization 
can be explained as a consequence of the dust alignment efficiency 
decreasing toward the inner parts of the dark clouds,
as all known mechanisms of grain alignment fail under the typical
physical conditions of dark cloud interiors.
On the other hand, beam smearing over small-scale field structure 
can also produce low polarization percentage (Rao et al.\ 1998); 
this can only be verified by observations with higher resolution.

Due to the apparent close anti-correlation of the polarization 
percentage and the total intensity in Figure 3, 
it is interesting to perform a least-squares linear fit
on these two quantities. 
We obtain $\log_{10}(p)=(-1.52\pm0.01)-(0.63\pm0.07)\times\log_{10}(I)$
with a correlation coefficient of $-$0.91 in W51 e1,
and $\log_{10}(p)=(-1.73\pm0.01)-(0.99\pm0.04)\times\log_{10}(I)$
with a correlation coefficient of $-$0.97 in W51 e2.
However, due to the difficulty of excluding the free-free 
emission at every position, direct comparison between 
our results and the dust alignment mechanisms cannot be done.
The tight correlations hint that the depolarization in the W51 e1/e2 cores 
is a gradual and global effect.  Therefore, it is unlikely that 
the depolarization is entirely caused by a sudden change of the alignment 
mechanism in a local region, as was reported in Orion-KL (Rao et al.\ 1998).

\section{Discussion}

\subsection{Magnetic Field Morphology}

The magnetic field direction inferred from the paramagnetic relaxation 
of grain alignment is perpendicular to the direction of the linear 
polarization of the dust emission (Davis \& Greenstein 1951).
Therefore, the average directions of the magnetic fields 
are 113\deg$\pm$5\deg\ in W51 e1 and 105\deg$\pm$7\deg\ in W51 e2 
(excluding e2 pol NW).
The field directions are approximately parallel to the minor axis 
of the two cores, suggesting that matter collapsed along the field 
lines to form the parent cloud of the double cores.   
The uniform field structure in the W51 e1/e2 cores also suggests that 
ionized gas in the UC \hii\ regions does not have much interaction 
with the magnetic field, or the interaction only exists 
on a scale smaller than our synthesized beam.

Comparison between the field morphology and the rotational
axes of W51 e1 and W51 e2 provides interesting implications
for the relation between magnetic fields and core formation.
Rotational motions have been observed toward these two cores 
by Zhang, Ho, \& Ohashi (1998) from analysis of the velocity
gradient of CH$_3$CN emission,
and the derived rotational axis is at 66\deg$\pm$27\deg\ 
for e1 and at 20\deg$\pm$17\deg\ for e2.  Clearly,
the rotational axes and the magnetic field directions 
of these two cores do not align with each other.
Although it has been suggested that protostellar disks 
may drag the field lines as they rotate (Holland et al.\ 1996),
it may not be the case for these two cores.
The field directions vary smoothly along the two cores.
Thus, it is unlikely that two independent rotating cores that 
twist the field lines separately would produce a matched field 
morphology.  Therefore, the smooth magnetic field structure is
most likely associated with the common envelope of e1 and e2.
Zhang, Ho, \& Ohashi (1998) caution that the uncertainty in their 
analysis may be large.   	However, if the difference 
between the directions of rotational axes and magnetic fields
is indeed significant, the magnetic field of the parent cloud 
seems to have had no effect in determining the axis of angular 
momentum of these two cores.

\subsection{Estimation of the Magnetic Field Strength}

Owing to the lack of understanding of the detailed mechanism 
of grain alignment, the magnetic field strength cannot be
directly derived from linear polarization measurements 
of dust emission (Lazarian, Goodman, \& Myers 1997). However,
Chandrasekhar \& Fermi (1953, hereafter CF) have proposed a method 
to estimate the field strength from the dispersion of polarization 
angles ($\delta\phi$) by assuming that the variation in polarization 
angles results from the perturbation of Alfv\'en waves on the field 
lines.  In this case, the field strength projected in the plane of 
the sky ($B_p$) is given by
\begin{equation}
B_p = Q \sqrt{4\pi\bar{\rho}}~~\frac{\delta v_{los}}{\delta\phi},
\end{equation}
\noindent where $\bar{\rho}$ is the average density, $\delta v_{los}$
is the rms line-of-sight velocity, and $Q$ is 1 for CF's derivation.
Detailed studies show that the CF formula tends to overestimate 
the field strength.  $Q$ can be reduced by several factors,
such as the inhomogeneity of clouds and the line-of-sight averaging 
(Zweibel 1990; Myers \& Goodman 1991).  Ostriker, Stone, \& Gammie 
(2001) calculate the value of $Q$ from their simulations
of turbulent clouds, and suggest that 
the CF formula with $Q\approx0.5$ can account for 
the complex magnetic field and density structure, and
provide accurate estimates of the plane-of-sky field strength
under strong field cases when $\delta\phi \leq 25\deg$.
Since the angle dispersion we measured in the W51 e1/e2 is small
($\delta\phi<9\deg$; \S3.2), it is interesting to work out
the field strength estimates with the modified CF formula.

However, the two other quantities, $\bar{\rho}$ and $\delta v_{los}$,
needed to complete this exercise have uncertainties not smaller 
than that of $\delta\phi$.  The density of the W51 e1/e2 cores varies 
with the physical components and scales, and we should use the average 
density obtained from the region that the observed polarization 
is associated with.
Zhang \& Ho (1997) measure the NH$_3$ emission in the infall regions of 
the cores and derive $n_{H_2}\sim3\times10^6\cm-3$,
which is probably too high for our purpose because the polarized intensity
in the W51 e1/e2 seems to originate in a more extended region.
From spectral decomposition of the whole W51\,A region 
(including W51\,e1/e2 and W51\,IRS 1), Sievers et al.\ (1991) show that 
W51\,A consists of a 20 K cold dust component with 
$n_H\sim5\times10^4\cm-3$ ($n_H=2n_{H_2}$) and a 60 K 
warm dust component with $n_H\sim10^6\cm-3$.  
This provides a range for the density 
as W51 e1/e2 may contain a mixture of both components.   
Because the column density of the warm dust is $\sim$5 times more
than that of the cold dust and W51\,e1/e2 is the densest region 
in W51\,A (Sievers et al.\ 1991), W51\,e1/e2 is more likely
to have a density close to $10^6\cm-3$.
It is also difficult to decompose $\delta v_{los}$
that is associated with Alfv\'enic motion
from the molecular linewidths which are broadened by the complicated
dynamics in W51 e1/e2, such as infall (3.5 \kms: Zhang et al.\ 1998), 
rotation ($\sim$4 \kms: Zhang \& Ho 1997), and possible outflow 
activities.   The linewidth of an optically thin NH$_3$ 
line measured in the envelope of e1 and e2 by Young, Keto, \& Ho (1998),
$\Delta v$=1.25 \kms, may contain little contamination from
dynamical motions.  Therefore, we adopt $\delta v_{los} = 
\Delta v / \sqrt{8\ln2} = 0.53 \kms$.

Given the large uncertainty of the input parameters, 
we can only calculate the lower limits of $B_p$
using the upper limits of the angle dispersions obtained in \S3.2.
The derived $B_p$ for the density limits of W51 e1/e2 
are listed in Table 2.  Our results show that the lowest value 
of $B_p$ in W51 e1/e2 is just comparable to the typical magnetic 
field strengths measured in molecular cores with $n_{H_2}=10^{5-6}$
\cm-3 using the Zeeman effect ($\sim500\mu$G: Crutcher 1999).   
Since this typical field strength provides crucial contribution
in cloud evolution (Crutcher 1999), our results show that 
the magnetic energy is an important component in the energetics
of the W51 e1/e2 envelope.
It can be shown that the ratio of the turbulent to magnetic energy 
is simply proportional to the square of the angle dispersion 
(Lai et al.\ 2001, in preparation).
The small angle dispersions we measured here imply that
the magnetic field dominates the turbulent motion 
in the regions that emit polarized flux in the W51 e1/e2 molecular cores.

\section{Conclusions}

We have measured linear polarization of the thermal dust emission 
at 1.3 mm toward the W51 e1/e2 cores with the BIMA array.
The conclusions of our observations are the following:

\begin{itemize}
\item The magnetic field is parallel to the minor axis of 
the W51 e1/e2 cores, consistent with the theoretical picture that 
cloud cores form due to gravitational contraction along field lines.
However, the significant difference between the directions
of magnetic field and rotational axes hints that the magnetic field
of the parent cloud does not determine the axis of the 
angular momentum of these two cores.

\item The polarization percentage decreases toward the high
intensity regions. The tight anti-correlation between $\log~p$
and $\log~I$ in both W51 e1 and W51 e2 
implies that depolarization is a global effect
and may be caused by the decreasing dust alignment efficiency
toward higher density regions.

\item The small dispersion of the polarization angles in W51 e1/e2
suggests that the magnetic field is strong ($\gtrsim$ 1 mG) and 
dominates turbulence in the regions where the polarized flux arises.
\end{itemize}

\acknowledgments

This research was supported by NSF grants AST 99-81363 and AST 98-20651.
We would like to thank the staff at Hat Creek, especially Rick Forster 
and Mark Warnock for assistance with the polarimeter control system.


\begin{table}
\caption{Polarization Measurements in the W51 e1/e2 regions}
\vspace*{0.3cm}
\begin{tabular}{lcccl}
\hline
\hline
Position & Stokes $I$ & Polarization & Polarization & Note \\
(\arcsec,\arcsec)\tablenotemark{a} & (\Jb) & Percentage(\%) & Angle (\deg) &\\
\hline

(-6.3,~3.3) & 0.25$\pm$0.03 & ~6.4$\pm$2.0 & 58$\pm$~8 & e2 pol NW	 \\ 
(-1.0,~2.0) & 0.35$\pm$0.03 & ~5.6$\pm$1.4 & ~1$\pm$~7 & e2  	 \\
(-3.6,~0.6) & 3.22$\pm$0.03 & ~0.7$\pm$0.2 & 19$\pm$~6 & e2 1mm peak \\
(-1.6,~0.3) & 0.91$\pm$0.03 & ~1.5$\pm$0.5 & 10$\pm$10 & e2	  \\
(-1.6,-2.9) & 0.32$\pm$0.03 & ~7.9$\pm$1.6 & 16$\pm$~5 & e2 south \\
(-3.9,-2.9) & 0.74$\pm$0.03 & ~3.1$\pm$0.6 & ~9$\pm$~6 & e2 south \\
(-2.0,-6.3) & 0.37$\pm$0.03 & ~6.5$\pm$1.4 & 19$\pm$~6 & e1	 \\
(-4.2,-6.3) & 1.79$\pm$0.03 & ~2.1$\pm$0.3 & 29$\pm$~4 & e1 1mm peak \\
(-4.2,-7.5) & 1.45$\pm$0.03 & ~3.0$\pm$0.3 & 28$\pm$~3 & Peak of polarized flux \\
(-6.5,-7.5) & 0.30$\pm$0.03 & ~5.8$\pm$1.6 & ~7$\pm$~8 & e1      \\
(-2.0,-9.5) & 0.20$\pm$0.03 & 10.0$\pm$2.7 & 26$\pm$~7 & e1      \\
(-4.2,-9.5) & 0.63$\pm$0.03 & ~4.8$\pm$0.8 & 28$\pm$~4 & e1	 \\
\hline
e1 average  & -- & ~3.2\e0.6 & 23\e~5 \\
e2 average  & -- & ~1.1\e0.3 & 15\e~7\tablenotemark{b}  \\
total average & -- & ~1.8\e0.4 & 21\e~6\tablenotemark{b} \\
\hline
\hline
\tablenotetext{a}{Offsets are measured with respect to
the phase center: $\alpha_{2000}$=19$^h$23$^m$44\fs2,
$\delta_{2000}$=14\deg30\arcmin33\farcs4.}
\tablenotetext{b}{e2 pol NW is excluded.}
\end{tabular}
\end{table}


\begin{table}
\caption{Estimates of magnetic field strengths in the plane of the sky}
\vspace*{0.3cm}
\begin{tabular}{lccccc}
\hline
\hline
Region & $\sigma_{\phi} (^{\circ})$\tablenotemark{a} 
& $\delta\phi_{obs} (^{\circ})$\tablenotemark{b} 
& $\delta\phi$ (\deg)\tablenotemark{c}
& $n_H$ (\cm-3)\tablenotemark{d}
& $B_p$ (mG)\tablenotemark{e} \\
\hline
e1   &4.9$\pm$1.4 & 4.8\e2.0 & $<5.8$ &
\begin{tabular}{@{}c@{}}
5\x10$^4$ \\
10$^6$
\end{tabular} & 
\begin{tabular}{@{}c@{}}
$>$ 0.3  \\
$>$ 1.3 
\end{tabular} \\
\hline
e2      & 6.7$\pm$3.4 & 6.6\e3.0 & $<9.0$ &
\begin{tabular}{@{}c@{}}
5\x10$^4$ \\
10$^6$
\end{tabular} & 
\begin{tabular}{@{}c@{}}
$>$ 0.2  \\
 $>$ 0.8
\end{tabular} \\
\hline
\hline
\tablenotetext{a}{The measurement uncertainty of the polarization angle}
\tablenotetext{b}{The observed polarization angle dispersion}
\tablenotetext{c}{The intrinsic polarization angle dispersion}
\tablenotetext{d}{The number density of atomic hydrogen}
\tablenotetext{e}{The plane-of-sky magnetic field strength}
\end{tabular}
\end{table}

\begin{figure}
\plotone{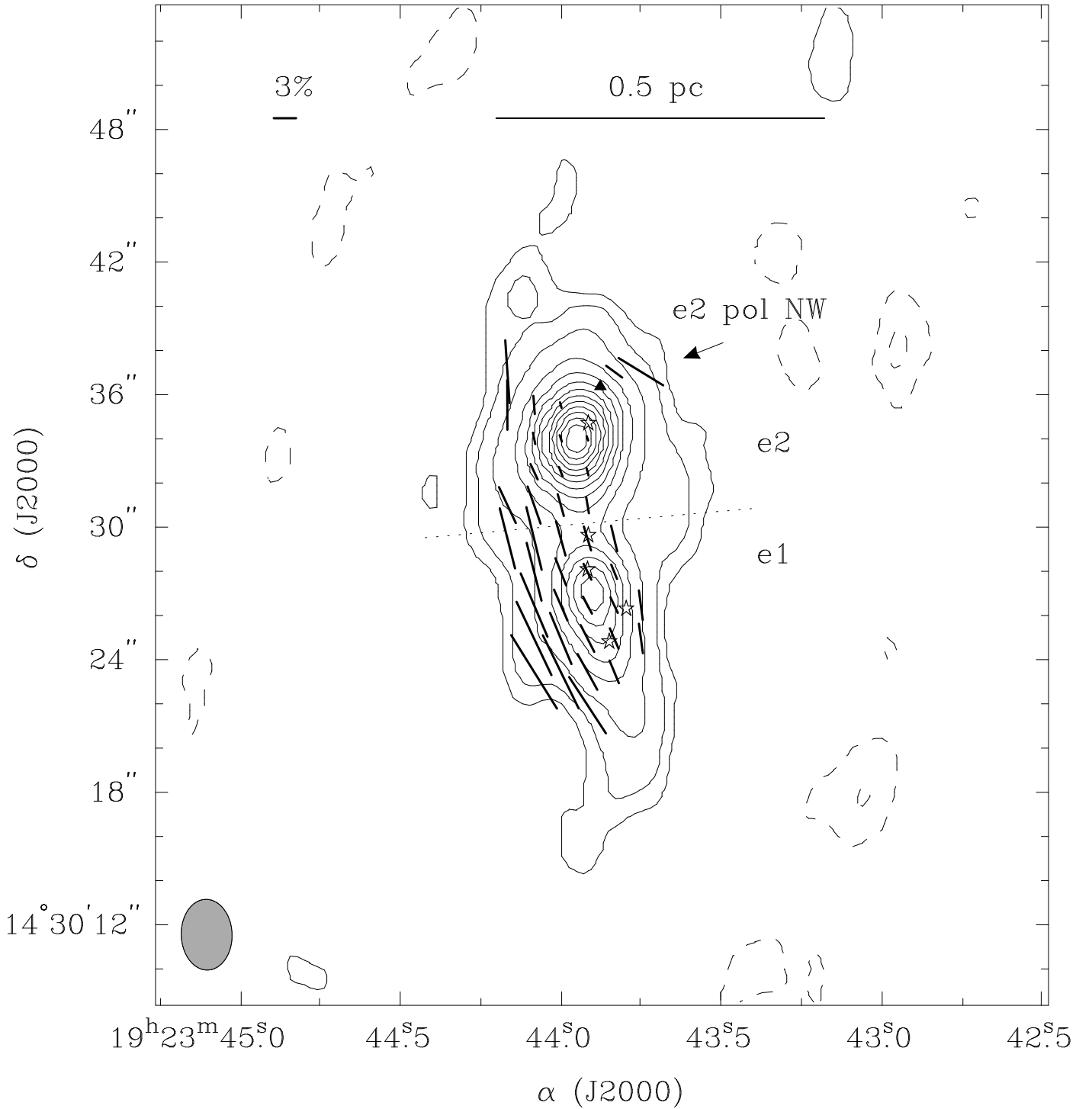}
\figcaption{Polarization map of the W51 e1/e2 cores.
The contours represent Stokes $I$ at $-5, -3, 3, 5, 10, 20, 30, ..., 
110~\sigma$ levels. The 1 $\sigma$ noise level of Stokes $I$ is 27 \mJb.
The line segments are polarization vectors, and
their lengths are proportional to the polarization percentage 
with a scale of 3\% per arcsec length.
The dashed line that crosses the saddle point of the continuum
is used to separate the data of W51 e1 and W51 e2 for our statistics.
The star symbols mark the positions of e2, e4, e8, e1 and e3 from north 
to south. The triangle marks the position of H$_2$O masers near e2 (W51 Main).
e2 pol NW refers to the two vectors northwest of W51 Main.
The ellipse indicates the synthesized beam of the Stokes $Q$ and $U$ images, 
which is 3\farcs2$\times$2\farcs3 with PA=2\deg. 
Positions with high intensity but without polarization detections 
may have higher fraction of free-free emission or/and 
have an environment inappropriate for the magnetic alignment 
of the dust grains.}
\end{figure}

\begin{figure}
\epsscale{0.65}
\plotone{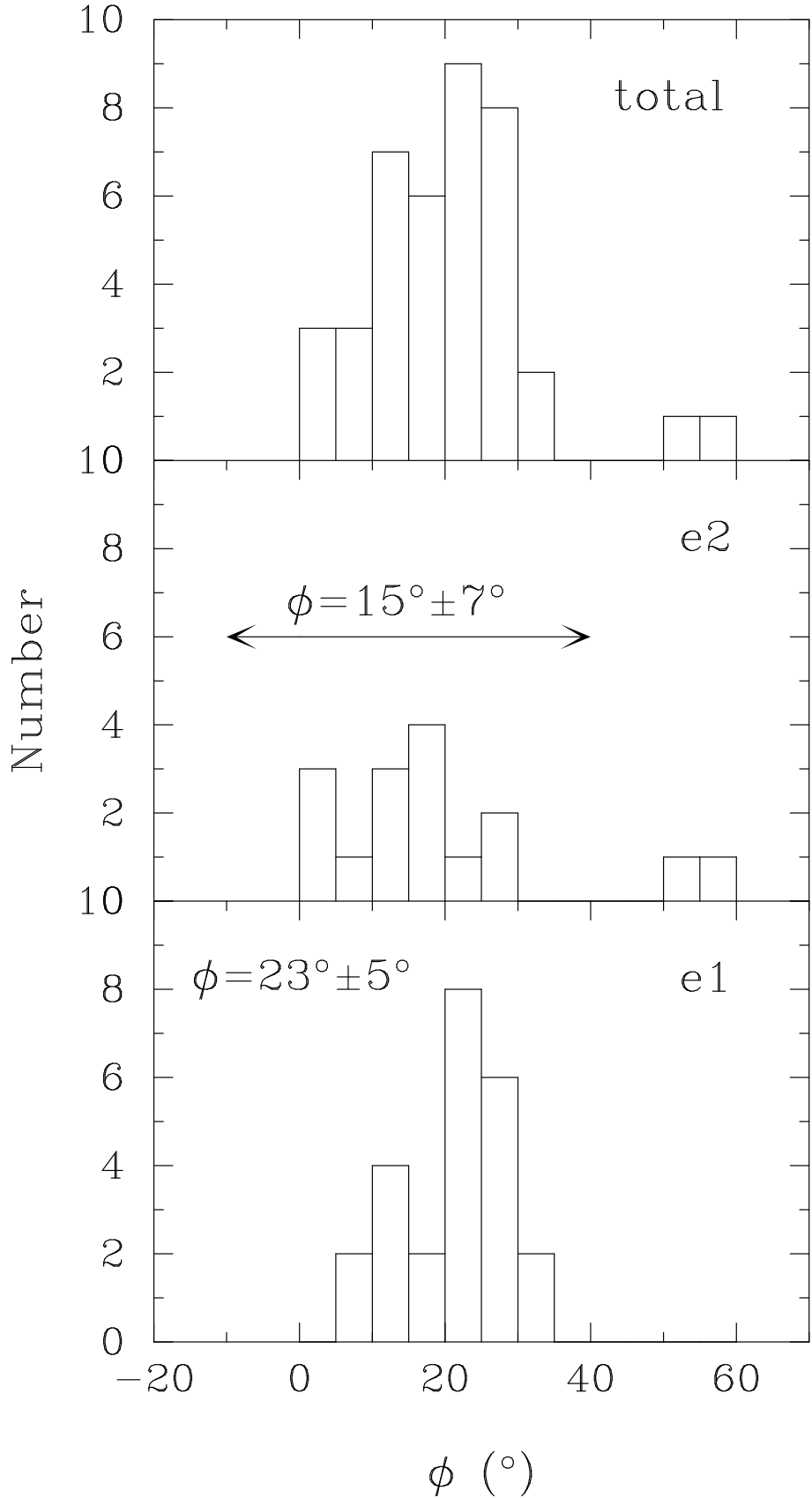}
\figcaption{Distribution of polarization angle in W51 e1/e2. 
The average and the dispersion of the polarization angle in e1 and e2 are
labeled, and the double arrow indicates that e2 pol NW with 
$\phi=50\deg-60\deg$ is excluded in the calculation.}
\end{figure}

\begin{figure}
\plotone{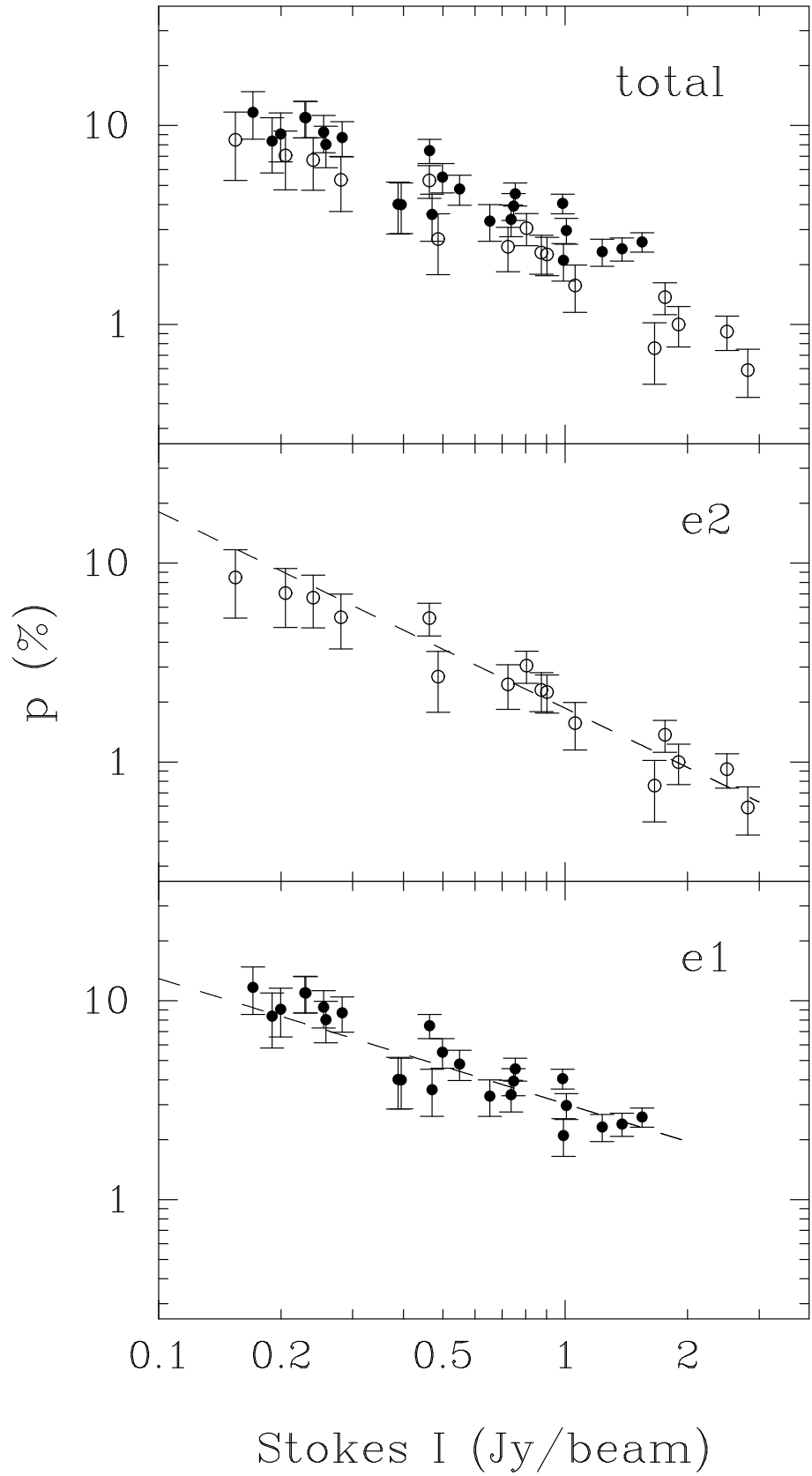}
\figcaption{ The variation of the polarization percentage with Stokes $I$ 
in W51 e1/e2.  The dashed lines are the results of the least-squares linear 
fit.}
\end{figure}

\end{document}